\documentclass[11pt]{article}

\usepackage{mons}
\usepackage{graphicx}
\usepackage{times}

\usepackage{amsmath}

\usepackage[mathcal]{eucal}

\newcommand{\astroncite} [2] {}

\newcommand{\eqna} [1] {
\begin{eqnarray} 
#1 
\end{eqnarray}}

\newcommand{\derivp} [2] {\frac {\partial #1 } {\partial #2} }

\def\note #1]{{\bf #1]}}

\def\plotfiddle#1#2#3#4#5#6#7{\centering \leavevmode
\vbox to#2{\rule{0pt}{#2}}
\includegraphics{#1}}

\begin{document}

\title{Oscillation power as a diagnostic tool for stellar turbulent spectra}

\author{R\'eza Samadi}
\affil{DESPA Observatoire de Paris-Meudon, F-92195 Meudon}

\author{G\"unter Houdek}
\affil{Institute of Astronomy, University of Cambridge, Cambridge CB3 0HA, UK} 

\begin{abstract}

\noindent Recent observations and theoretical studies support the theory that
solar-type oscillations are intrinsically stable but excited
stochastically by the turbulent convection in the outer layers of the
star. The acoustic noise generated by the convective motion depends
on the details of the turbulent energy spectrum. In this paper we
present a general formulation for the acoustic noise generation rate based
on previous works by other authors.  In this general formulation any model
for the spatial turbulent energy spectrum and for the turbulent time spectrum
can be assumed.  We compute acoustic power spectra and oscillation amplitudes
of radial oscillations for models of the Sun and Procyon A. The results are
compared with recent observations. 

\end{abstract}

\section{Introduction}
\noindent Instruments aboard the SOHO spacecraft have provided high-quality 
data of solar oscillations. Comparing theoretical estimates of acoustic power
spectra with such high-quality data leads to a better understanding of
the excitation processes of p-mode oscillations and provides more details
of the characteristics of stellar turbulence. The acoustic power injected 
into the p modes by turbulent convection has been modelled by several 
authors (e.g., Goldreich \& Keeley 1977; 
Christensen-Dalsgaard \& Frandsen 1983; 
Balmforth 1992b; Goldreich, Murray \& Kumar 1994; Musielak et al. 1994).
Balmforth (1992b) and Goldreich, Murray \& Kumar (1994) investigated the
contributions of the fluctuating entropy to the noise generation rate
additional
to the contributions of the fluctuating Reynolds stresses but reported
different results. Balmforth found the entropy fluctuations to be less
important relative to the Reynolds stress contributions. Goldreich, Murray
\& Kumar (1994, hereafter GMK), however, concluded that the fluctuating
entropy contribution is about 3--4 times larger than the Reynolds stress
contribution, a result which was also found by the hydrodynamical simulations
of Stein \& Nordlund (1991).

Here, we adopt the formulation of Samadi et~al. (2000), which takes into account
contributions both from the Reynolds stresses and from the fluctuations of
the entropy.
Moreover, the new formulation allows a consistent investigation of the 
effects of using different forms of the turbulent time spectrum and turbulent
energy spectrum.  In particular, we study the effects of using various forms
of the spatial turbulent energy spectrum and turbulent time spectrum on the
acoustic power of radial p-mode oscillations for a model of the Sun and for
a model of Procyon A and compare the results with recent observations.
Better agreement with the observations is found for calculations in which
the turbulent energy spectrum includes contributions from convective elements
with spatial scales larger than one mixing length.


\vspace{-7mm}
\section{Excitation of stellar p modes by turbulent convection}

\subsection{Acoustic noise generation rate} 

\noindent The acoustic power injected into the oscillations is defined 
(e.g., GMK) in terms of the damping rate $\eta$, the mean-square amplitude 
$\langle A^2 \rangle$, the mode inertia $I$ and oscillation frequency 
$\omega_0$: 
\eqna{
P = {2\,\eta}\;1/2{\langle A^2 \rangle}\;I\;\omega_0^2\,.
}
The mean-square amplitude is determined
by the balance between the energy gain from the turbulent flow and the 
energy drain by thermal and mechanical damping processes. It can be 
derived as (e.g., Balmforth 1992b)
\eqna{
\begin{array}{ccc}
\left < A^2 \right > \propto\eta^{-1}\int_{0}^{M}dm\;\rho_0\;w^4 \;\left
(\derivp { \xi_r} {r} \right )^2  \;   \mathcal{S}(\omega_0,m)  & \textrm{with}
&
\mathcal{S}(\omega_0,m) = \int_0^\infty dk \; \phi( k)\;\chi_k (\omega_0)\,,
\end{array}
\label{eqn:A2}
}
where $\displaystyle{\xi_r }$ is the radial displacement eigenfunction, 
$\rho_0$ the density, $k$ the wavenumber of an eddy, $w$ is the vertical 
rms velocity of the convective elements and $\mathcal{S}$ is the turbulent 
source term, describing contributions to the noise generation rate from 
eddies with different sizes.
The integration is performed over the stellar mass $M$.  
The mean-square amplitude is indirect proportional to the damping rate, $\eta$, 
(e.g. Balmforth 1992b) and consequently the noise generation rate 
(or acoustic power) $P$ becomes independent of $\eta$. Thus comparing
the estimated noise generation rates with observations avoids the additional
uncertainties in the modelling of the damping rates (GMK).
Both Reynolds stresses and entropy fluctuations contribute to the excitation
of p modes, and proper models are needed for their computations.
The turbulent source $\mathcal{S}$, describing the turbulent spectrum, can 
be separated into a spatial turbulent energy spectrum $\phi(k)$ and in a 
turbulent time spectrum $\chi_k(\omega)$ (Stein 1967). 
The present formulation differs from the formulation of Balmforth (1992b) 
and GMK mainly in the way the entropy fluctuations are modelled 
(for more details see Samadi et~al. 2000).

\vspace{-7mm}
\subsection{Turbulent ingredients}
\noindent The correlation time of the turbulent eddies is modelled
by the turbulent time spectrum $\chi_k(\omega)$; several forms of 
$\chi_k(\omega)$ have been proposed: 
the commonly used Gaussian spectrum (e.g., Goldreich \& Keeley 1977, 
Balmforth 1992b), the exponential spectrum (Stein 1967; Musielak et~al. 1994) 
and the ``Modified Gaussian'' spectrum as suggested by Musielak et~al. (1994).  
Similarly, various forms for the spatial energy spectrum $\phi(k)$ have 
been proposed:
the Kolmogorov spectrum (KS), the spectrum suggested by Spiegel (1962)
which includes a representation for the nonlinear interaction between 
turbulent modes (SS), and the ``Raised Kolmogorov'' spectrum (RKS) 
proposed by Musielak et~al. (1994), which was derived empirically from 
the solar observations of Muller (1989).
Only the RKS spectrum includes a description that takes into account
contributions from eddies having smaller wavenumbers than those considered
in the KS and SS spectra. These eddies correspond to meso-granulation. 

\section{Solar model and calibration}

\noindent We considered a solar model computed with the CESAM code (Morel 1997)
assuming the model parameters displayed in Fig.~1.
The oscillation properties were obtained from the adiabatic FILOU pulsation 
code of Tran Minh \& Leon (1995). 
We computed models with various time spectra and concluded that
the Gaussian time spectrum provides the best agreement between
the computed and observed (Libbrecht 1988) acoustic power spectrum. In 
particular the shape of the acoustic spectrum and the frequency of the 
maximum value of the acoustic power are closest to the observations for
models computed with the Gaussian time spectrum.

Using the observed linewidths of Libbrecht (1988) we computed the mean 
surface velocities from the power estimates. The results are plotted
in Fig.~1 as functions of oscillation frequencies 
for model computations using the RKS (continuous curve), the KS (dashed curve)
and the SS (dot-dashed curve) spectrum. For all three model computations
the Gaussian time spectrum was assumed.
Amplitudes obtained with the RKS spectrum are closest to the observed values. 
Moreover, use of the RKS spectrum leads to the smallest frequency shift
between the computed and observed maximum value of the velocity amplitudes.

In order to estimate acoustic power spectra for other stars we need to
calibrate our formulation, i.e., to scale the free parameters, which are
inherent in the formulation of the noise generation rate (for details see
Samadi et~al. 2000). The noise generation rate is calibrated for all three 
turbulent spectra in such a way as to predict the same maximum value for 
the velocity amplitude of $18\,{\rm cm\,s}^{-1}$ as suggested by the
BBSO observations of Libbrecht (1988).

\begin{figure}[t]
\plotfiddle{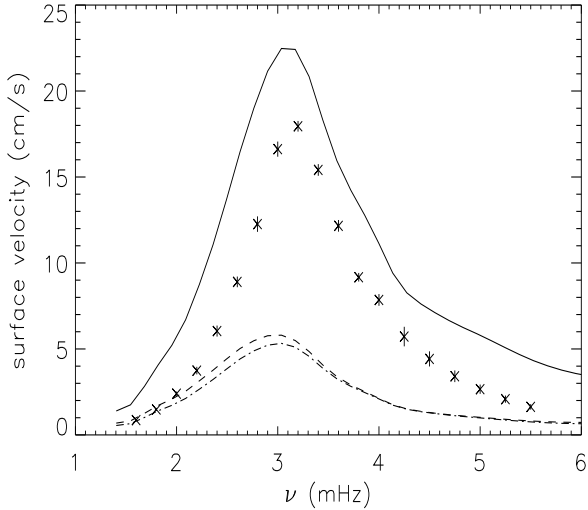}{5.0truecm}{0}{50}{60}{-243pt}{-50pt}
\parbox{\textwidth}{\hspace{8.5cm}
   \parbox{7.9cm}{
      \vspace{-60truemm}
      \caption{\footnotesize
               Computed surface velocities for the Sun obtained from 
               computations including the contributions from both the Reynolds 
               stresses and the entropy fluctuations, and assuming different
               spatial turbulent energy spectra: the continuous curve 
               displays the results for the RKS spectrum, the dashed curve 
               for the KS and the dot-dashed curve for the SS spectrum.
               The Gaussian time spectrum is used in all three cases.
               Crosses represent the solar measurements of Libbrecht (1988).
               The solar model has an age of 4.6 Gy and its composition
               for the abundance by mass of hydrogen and heavy elements 
               is $Y$\,=\,0.2682, $Z$\,=\,0.0175, respectively. For the 
               mixing-length parameter a value of $\alpha\,=\,1.785$ is used 
               and the effective temperature 
               \hbox{$T_{\rm eff}$\,=\,5782\,K}.
              }
   }
}
\label{fig:VRSEP_spc}
\end{figure}

\vspace{-5mm}
\section{What can we learn from other stars ?}

\noindent Using the same programmes with the same input physics as used for 
the solar model described above, we computed a model for Procyon A. For the 
model parameters we assumed a mass $M\,=\,1.46\, M_{\odot}$, an effective 
temperature $T_{\rm eff}\,=\, 6395\,$K, $\alpha\,=\,1.785$ and the same 
chemical composition as for the solar model (see Fig.~1).
The left panel of Fig.~2 shows the normalized power versus the oscillation 
frequencies computed with the RKS (continuous curve), the KS (dotted curve) 
and the SS (dot-dashed curve) spectrum. The results suggest large differences 
at high frequencies between models computed with the RKS and the KS spectrum.
At low frequencies the shape of the noise generation rate (power) is
predominantly determined by the modal inertia, whereas at high frequencies
the shape of the eigenfunctions becomes more important. This dependence
on the eigenfunctions at high frequencies is more pronounced in the model
of Procyon than in the solar model.
However, only small differences are found between power estimates 
obtained from computations including only the Reynolds stress contribution 
and for computations including both Reynolds stress and entropy 
contributions, assuming the same turbulent spectrum.

In order to compute velocity amplitudes we need estimates of the pulsation
damping rates, $\eta$. The damping rates for radial oscillations were obtained 
from a non-adiabatic pulsation programme introduced by Balmforth (1992a). 
In this programme convection is treated with a time-dependent, nonlocal 
generalization of the mixing-length formulation (Gough 1976, 1977). 
Computation details can be found in Balmforth (1992b) 
and in Houdek et~al. (1999). The right panel of Fig.~2 shows the estimated 
surface velocities computed with the RKS and the KS spectra assuming 
the computed damping rates $\eta$ and the same parameters as suggested by 
a solar model which has been calibrated to the observations.
We observe a large frequency shift of the maximum values of the estimated
velocities between models computed with the RKS and the KS spectrum. 

\begin{figure}[t]
\plotfiddle{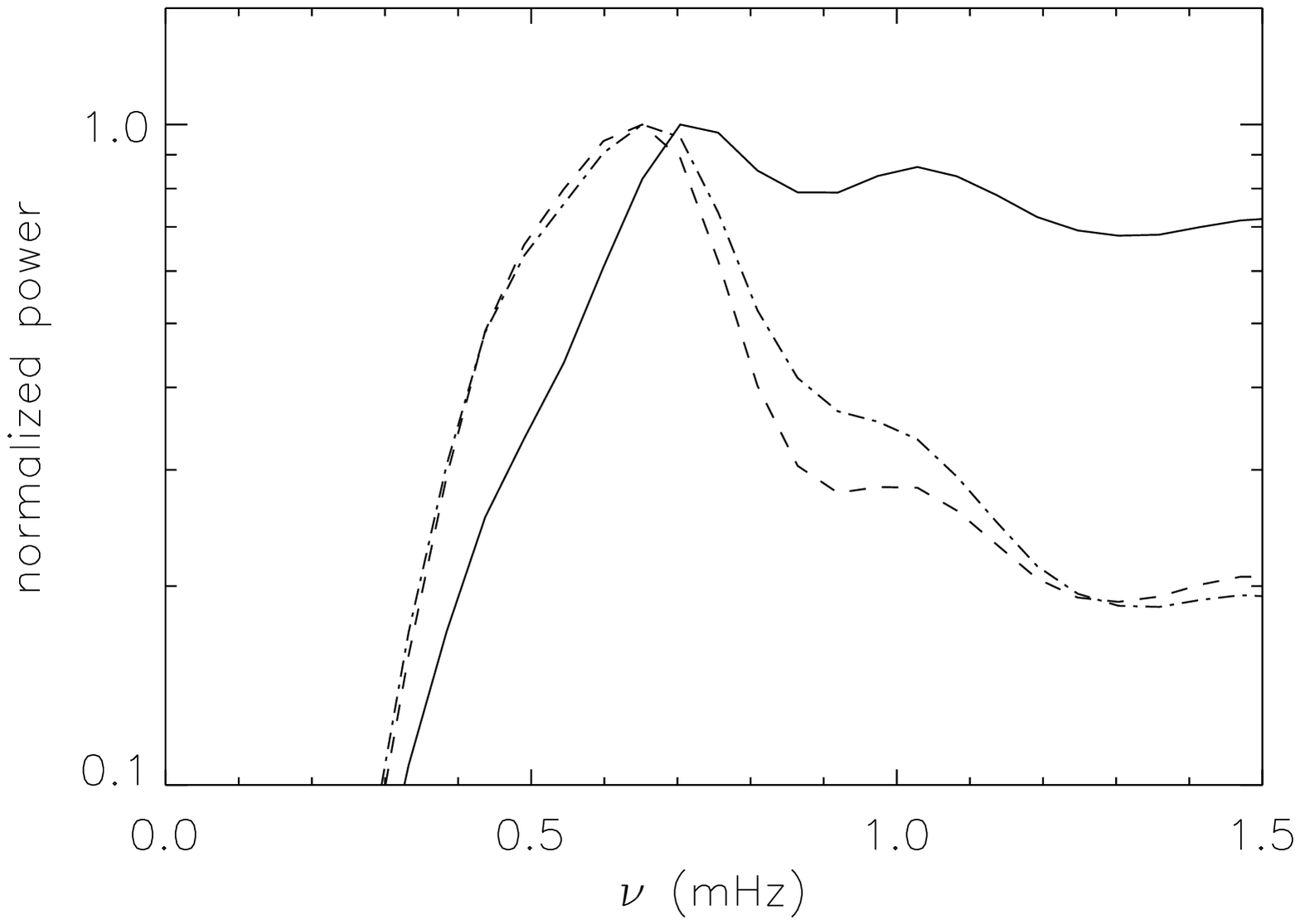}{2.9truecm}{0}{47}{60}{-243pt}{-112pt}
\plotfiddle{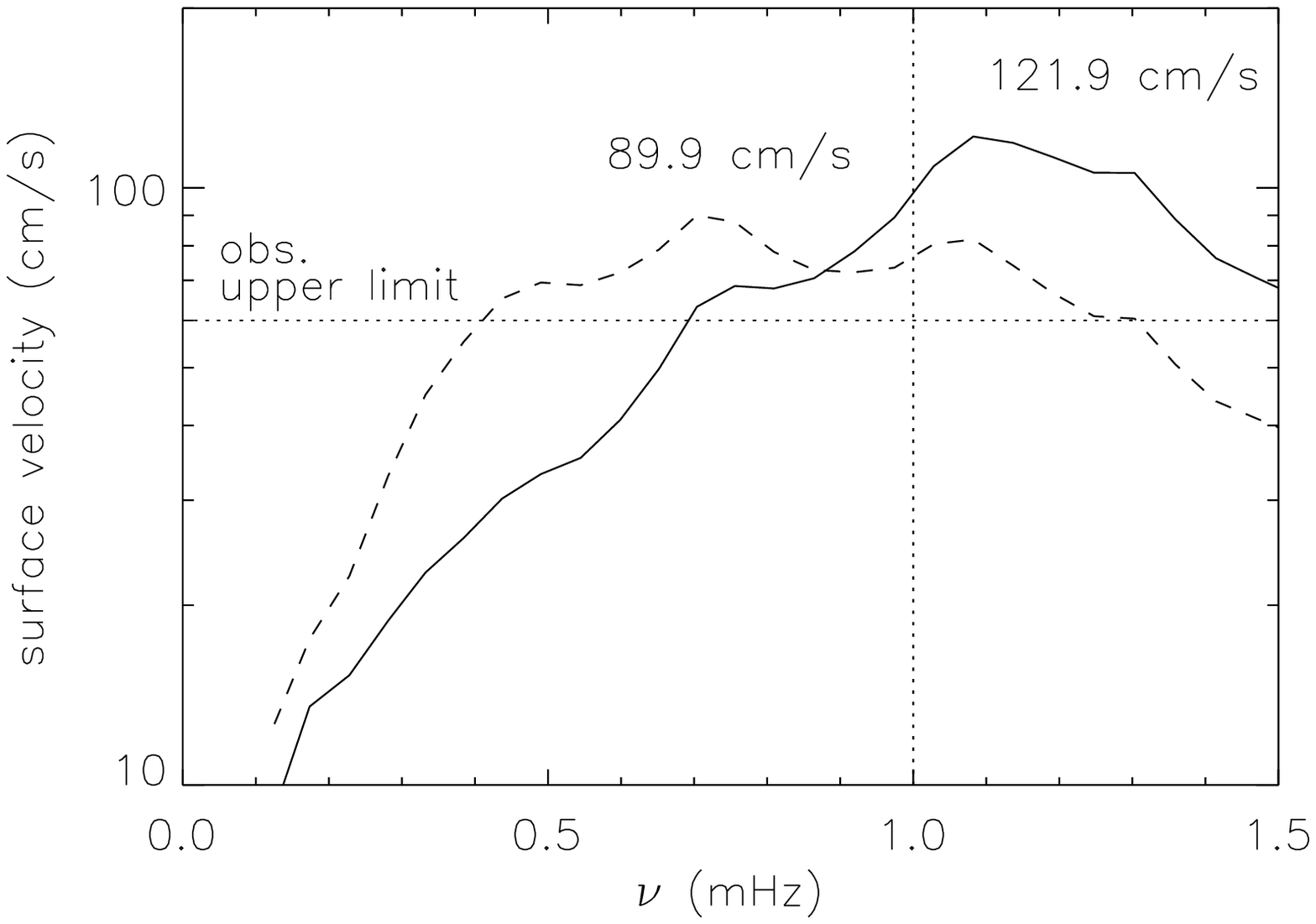}{2.9truecm}{0}{47}{60}{-1pt}{-14pt}
\caption{\footnotesize
           {\bf Left panel}: Acoustic power estimates of p-mode oscillations 
            for a model of Procyon A. The results are obtained from 
            computations including the contributions from both the Reynolds 
            stresses and the entropy fluctuations, and assuming different
            turbulent spectra: the RKS (continuous curve), the KS (dashed 
            curve) and the SS (dot-dashed curve) turbulent spectra.
           {\bf Right panel}: Expected surface velocities for a model of 
            Procyon A. The continuous curve displays the results obtained with
            the RKS spectrum and the dashed line shows the results obtained with
            the KS spectrum. The observational upper limit of $\approx 60\,{\rm 
            cm\,s}^{-1}$ is indicated by the straight horizontal dotted
            line. The straight vertical dotted line indicates the position 
            of frequency of the maximum value of the observed velocities.
           }
\vspace{-2mm}
\label{fig:pRSEPpro_spc}
\end{figure}

In the calculations for the amplitudes we assumed that the effect of
using different eigenfunctions for computing the power with the
programme of Tran Minh \& Leon and for the damping rates with the programme
of Balmforth are small compared to the uncertainties inherent in the 
formulation of estimating the acoustic power. This inconsistency may affect 
the absolute value of the surface velocity but has no effect on the comparison 
between velocity estimates obtained with different turbulent spectra.

Observations of Procyon have been carried out by Martic et~al. (1999) 
and Barban et~al. (1999). These authors concluded that the maximum velocity 
is observed around  $\nu \simeq 1 \, {\rm mHz}$ with an upper limit 
of $V_{\rm max} \lesssim 60 \, {\rm cm\,s}^{-1}$. The velocity
estimates using the RKS spectrum (continuous curve in the right panel of
Fig.~2) exhibit a maximum at $\nu \simeq 1 \, {\rm mHz}$, in fair agreement 
with the observations. However, the predicted surface velocities are
too large relative to the observations. It should be noted that some of the 
uncertainties in the computed amplitudes do stem from the uncertainties in 
the damping rate estimations, $\eta$.

\section{Conclusion and perspectives}

\noindent We have presented results of a more general and consistent 
formulation for estimating the acoustic noise generation rate in 
solar-type stars. In accordance with GMK, the entropy contribution has 
been found to be roughly three times larger than the contributions from 
the Reynolds stresses. However, preliminary results suggest only small
differences in the maximum amplitude values in other stars between
models computed with the Reynolds stress contribution alone and models
computed with both Reynolds stress and entropy contributions, provided
the amplitudes in both cases are calibrated first to solar observations 
for a solar model.

For the solar case we conclude that the ``Raised Kolmogorov'' spectrum  
(RKS) and the Gaussian time spectra provide the best agreement with the 
observations. Computations for a model of Procyon A support this 
conclusion. For the Procyon model the differences between the RKS and 
the KS spectra are larger than in the solar case.

Further investigations of the proposed formulation are necessary such as 
testing it against the results of hydrodynamical simulations. Extending 
the investigation of the noise generation rate to other stars improves 
our understanding of stellar turbulence. Moreover, the results of modelling 
solar-type oscillation properties are of great importance for the selection
process of target stars in future space projects such as COROT (COnvection 
and ROTation), MONS (Measuring Oscillations in Nearby Stars) or MOST 
(Microvariability \& Oscillations of STars).

\vspace{-8mm}
\subsection*{Acknowledgments}
\noindent We are grateful to Y. Lebreton for the computation of models. We 
thank F. Tran Minh and L. Leon for the use of the FILOU pulsation code,
and we are indebted to M.J. Goupil for useful discussions.

\vspace{-5mm}

\baselineskip=0.85\normalbaselineskip

\vspace{-5mm}
\baselineskip=\normalbaselineskip
\discussion

DOUGLAS GOUGH  : Why does the ``Raised Kolmogorov'' spectrum lead to 
larger power estimates at high frequencies ?

\medskip

REZA SAMADI : The  ``Raised Kolmogorov''  spectrum (RKS) leads to a smaller
depth of excitation. The RKS takes into account eddies of smaller wavenumbers 
than the Kolmogorov spectrum (KS). There is also an excess of power at low
wavenumbers. Consequently, for a given wavenumber k the correlation time of 
an eddy ($\tau_k$) is larger with the RKS than with the KS spectrum. The major 
contribution to mode excitation comes from eddies with $\tau_k \omega_0 
\lesssim 1$. Thus the region where $\tau_k \omega_0 \lesssim 1$ should 
also be smaller with the RKS than with the KS. Therefore the RKS induces 
a smaller depth of the excitation region.


\begin{thebibliography}{}

\bibitem[\protect\astroncite{{Balmforth}}{1992}]{Balmforth92a}
{Balmforth}, N.~J., 1992a,
\newblock {\mnras} {255}, 603

\bibitem[\protect\astroncite{{Balmforth}}{1992}]{Balmforth92c}
{Balmforth}, N.~J., 1992b,
\newblock {\mnras} {255}, 639

\bibitem[\protect\astroncite{{Barban} et~al.}{1999}]{Barban99}
{Barban}, C., {Michel}, E., {Martic}, M., {Schmitt}, J., {Lebrun}, J.~C.,
  {Baglin}, A., and {Bertaux}, J.~L., 1999,
\newblock {\aap} {350}, 617

\bibitem[\protect\astroncite{{Christensen-Dalsgaard}and{Frandsen}}{1983}]{DF83}
{Christensen-Dalsgaard}, J., and {Frandsen}, S., 1983,
\newblock {Solar Physics} {82}, 489

\bibitem[\protect\astroncite{{Goldreich} and {Keeley}}{1977}]{GK77}
{Goldreich}, P., and {Keeley}, D.~A., 1977,
\newblock {\apj} {212}, 243

\bibitem[\protect\astroncite{{Goldreich} et~al.}{1994}]{GMK94}
{Goldreich}, P., {Murray}, N., and {Kumar}, P., 1994,
\newblock {\apj} {424}, 466


\bibitem[\protect\astroncite{Gough}{1976}]{DOG76}
{Gough}, D.~O., 1976,
\newblock in: {Problems of stellar convection},
     E. {Spiegel},J.-P. {Zahn} (eds), Springer-Verlag, Berlin, p. 15


\bibitem[\protect\astroncite{Gough}{1977}]{DOG77}
{Gough}, D.~O., 1977,
\newblock {\apj} {214}, 196

\bibitem[\protect\astroncite{{Houdek} et~al.}{1999}]{Houdek99}
{Houdek}, G., {Balmforth}, N.~J., {Christensen-Dalsgaard}, J., and {Gough},
  D.~O., 1999,
\newblock {\aap} {351}, 582

\bibitem[\protect\astroncite{{Libbrecht}}{1988}]{Libbrecht88}
{Libbrecht}, K.~G., 1988,
\newblock {\apj} {334}, 510

\bibitem[\protect\astroncite{{Martic} et~al.}{1999}]{Martic99}
{Martic}, M., {Schmitt}, J., {Lebrun}, J.-C., {Barban}, C., {Connes}, P.,
  {Bouchy}, F., {Michel}, E., {Baglin}, A., {Appourchaux}, T., and {Bertaux},
  J.-L., 1999,
\newblock {\aap} {351}, 993

\bibitem[\protect\astroncite{{Morel}}{1997}]{Morel97}
{Morel}, P., 1997,
\newblock {\aaps} {124}, 597

\bibitem[\protect\astroncite{{Muller}}{1989}]{Muller89}
{Muller}, R., 1989,
\newblock in: Solar and Stellar Granulation, R. {Rutten} and G. {Severino} 
          (eds.), Kluwer Academic Publishers, p. 101

\bibitem[\protect\astroncite{{Musielak} et~al.}{1994}]{Musielak94}
{Musielak}, Z.~E., {Rosner}, R., {Stein}, R.~F., and {Ulmschneider}, P., 1994,
\newblock {\apj} {423}, 474





\bibitem[\protect\astroncite{{Samadi}}{2000}]{Samadi2000}
{Samadi},R., {Goupil},M.-J., and {Mangeney},A., 2000,
\newblock {in preparation}

\bibitem[\protect\astroncite{{Spiegel}}{1962}]{Spiegel62}
{Spiegel}, E., 1962,
\newblock {J. Geophys. Res.} {67}, 3063

\bibitem[\protect\astroncite{{Stein}}{1967}]{Stein67}
{Stein}, R.~F., 1967,
\newblock {Solar Physics} {2}, 385

\bibitem[\protect\astroncite{{Stein} and {Nordlund}}{1991}]{SteinNo91}
{Stein}, R.~F., and {Nordlund}, {\AA}., 1991,
\newblock in: in: Challenges to Theories of the
     Structure of Moderate Mass Stars, D.O. {Gough} and J. {Toomre} (eds.),
     Springer-Verlag, Heidelberg, p. 195

\bibitem[\protect\astroncite{{Tran Minh} and {Leon}}{1995}]{Tran95}
{Tran Minh}, F., and {Leon}, L., 1995,
\newblock in {Physical Process in Astrophysics}, I.W. {Roxburgh} and 
       J.-L. {Masnou} (eds.), Springer-Verlag, Berlin, p. 219
\end{thebibliography}
\end{document}